\documentclass[preprint]{revtex4}

\usepackage[centertags]{amsmath}
\usepackage{graphicx}

\begin{document}

\title{Shot noise in an electron waveguide $\sqrt{NOT}$ gate }
\author{Linda E. Reichl and Michael G. Snyder \\
Center for Studies in Statistical Mechanics and Complex Systems,\\
The University of Texas at Austin, Austin, Texas 78712\\}

\begin{abstract}

We present a calculation of the shot noise in a ballistic electron waveguide $\sqrt{NOT}$ gate.  A general expression for the shot noise in the leads connected to these types of gates is shown.  We then parameterize an $S$-matrix which qualitatively describes the action of a $\sqrt{NOT}$ gate previously found through numerical methods for $GaAs/Al_xGa_{1-x}As$ based waveguides systems.  Using this $S$-matrix, the shot noise in a single output lead and across two output leads is calculated.  We find that the  measurement of the shot noise across  two output leads allows for the determination of the fidelity of the gate itself.

\end{abstract}

\maketitle

\section{Introduction}

A quantum computer constructed from ballistic electron waveguides consists of network of dual-waveguide qubits, $\sqrt{NOT}$ gates and entangling regions \cite{kn:ion}, \cite{kn:snyder}.  The network is hard wired and is constructed to perform specific computations.  A computation begins by injecting electrons into the input end of the device where they subsequently travel towards the output end of the device encountering $\sqrt{NOT}$ gates and entangling regions along the way.  The electrons emerge on the output end of the device and the computation is completed.     

An electron waveguide qubit consists of two parallel waveguides.  An electron in one waveguide  corresponds to a qubit state $|0\rangle$ while an electron in the other waveguide corresponds to a qubit state  $|1\rangle$.  As was shown in 
\cite{kn:akguc} a $\sqrt{NOT}$ gate can be implemented in a $GaAs/Al_xGa_{1-x}As$ based system of electron waveguides by creating a cavity connecting the waveguides. The cavity  consists of a region of reduced potential between the two waveguides of a single qubit and allows the tunneling of the electron between the waveguides.  A correct choice of the parameters of the cavity region creates an "electron beam splitter", so that a qubit initially in either the $|0\rangle$ or $|1\rangle$ state is in an equal superposition of the two after the electron passes through the cavity.  

It has been shown that the fidelity of a computation in these waveguide networks is strongly influenced by reflection of electron probability at the $\sqrt{NOT}$ gates \cite{kn:akguc}, \cite{kn:snyder}.  It is therefore essential to characterize the reflection properties of these $\sqrt{NOT}$ gate regions in a manner that is experimentally measurable.  The equal splitting of electron probability into each of the outgoing waveguides does not guarantee a high fidelity gate, as some or most of the electron probability could still be reflected back to the input side of the gate.  Below we show that the shot noise inherent in a $\sqrt{NOT}$ gate can provide another means for determining the quality of a particular $\sqrt{NOT}$ implementation.

Shot noise in mesoscopic semi-conductor wires has been studied extensively \cite{kn:landauer}, \cite{kn:blanter}.   The noise itself is due to the quantization of electron charge and is seen at very low temperatures when the thermal noise is small.  In electron waveguides, ballistic leads are connected to large reservoir regions where electrons can enter the leads without being scattered.  The leads are then connected to a scattering region, where the scattering amplitudes for the scattering region are known.  The shot noise measured in the current in the leads then can be expressed in terms of the scattering amplitudes and the distribution functions that describe the reservoirs.

Below we first describe the $\sqrt{NOT}$ gate in terms of leads, reservoirs and a scattering region.  We then use a scattering analysis to find the zero frequency shot noise in the leads connected to the $\sqrt{NOT}$ gate.  Once the shot noise for a general $\sqrt{NOT}$ gate is presented, a specific gate is studied which represents the scattering properties found in the numerically modeled gate in \cite{kn:akguc}.  The shot noise is shown to be a very good indicator of the overall fidelity of a physical $\sqrt{NOT}$ gate in electron waveguide qubits.

\section{Shot noise theory}

A ballistic electron waveguide qubit consists of two parallel waveguides coupled by cavities which allow flow of electron probability amplitude between the two waveguides which form the qubit.  The distribution  of the electron probability amplitude between the two  waveguides represents the state of the qubit.  A $\sqrt{NOT}$ gate is a cavity which connects the waveguides and through which the electron can tunnel to create an equal superposition of states $|0\rangle$ and $|1\rangle$.  We will follow a scattering approach to describe the shot noise associated with a ballistic electron waveguide $\sqrt{NOT}$ gate \cite{kn:blanter}.

\begin{figure}[htb]
\begin{center}
\includegraphics{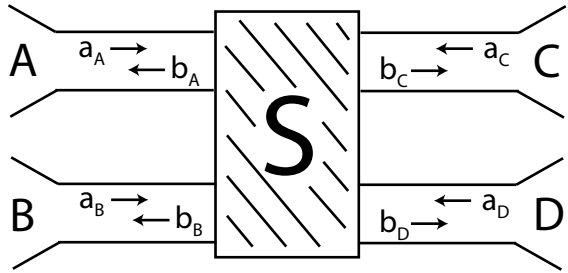}
\caption{A schematic of the a $\sqrt{NOT}$ gate.  Four leads $A$, $B$, $C$, and $D$ are connected to the scattering region where electron probability may tunnel between the waveguides.  Electrons in the upper two leads, $A$ or $C$, correspond to a qubit state of $|1\rangle$.  Electrons in the lower two leads, $B$ or $D$, correspond to a qubit state of $|0\rangle$.   The operators $\hat{a}_\alpha$ annihilate electrons in the incoming state in lead $\alpha$.  The operators $\hat{b}_\alpha$ annihilate electrons in the outgoing state in lead $\alpha$.}\label{gate}
\end{center}
\end{figure}

We consider a $\sqrt{NOT}$ gate consisting of four straight leads, $A$, $B$, $C$,  and $D$ connected to a scattering region where an incoming electron may be scattered into any of the four leads (Fig. \ref{gate}).  

An electron traveling in one of the leads propagates freely and can be entering or leaving the scattering region.  Each of the four leads is connected to a separate reservoir of electrons at temperature $T$ and distributed according to the  Fermi distribution function $f_\alpha=[\mbox{exp}[k_B\beta(E-\mu_\alpha)+1]^{-1}$, where $\mu_\alpha$ is the chemical potential of reservoir $\alpha=A,B,C,D$.  Current is driven through the system by biasing one or more of the reservoirs at a small negative potential relative to the other reservoirs.

We assume the energy of the incoming electrons in the system can be controlled such that only the first transverse mode is occupied in the waveguide and we neglect any scattering of electrons into other transverse modes.  The electron waveguides then can be treated as quasi-one-dimensional quantum wires.  We use creation and annihilation operators to describe electrons in scattering states in the leads.  Operators $\hat{a}^\dagger_\alpha(E)$ and $\hat{a}_\alpha(E)$ create and annihilate {\it{incoming}} electrons in lead $\alpha$.  Operators $\hat{b}^\dagger_\alpha(E)$ and $\hat{b}_\alpha(E)$ create and annihilate {\it{outgoing}} electrons in lead $\alpha$.  The operators obey the commutation relation,
\begin{equation}
\hat{a}^\dagger_\alpha(E)\hat{a}_\beta(E')+\hat{a}_\beta(E')\hat{a}^\dagger_\alpha(E)=\delta_{\alpha\beta}\delta(E-E'),
\end{equation}

The operators for incoming states are related to the operators for the outgoing states by a scattering matrix, $S$, so that
\begin{equation}
\left(\begin{array}{r}\hat{b}_{A}\\ \hat{b}_{B}\\ \hat{b}_{C}\\\hat{b}_{D}\end{array}\right) = S\left(\begin{array}{r}\hat{a}_{A}\\ \hat{a}_{B}\\ \hat{a}_{C}\\ \hat{a}_{D}\end{array}\right)
\end{equation}
where
\begin{equation}
S = \left(\begin{array}{rrrr}r_{AA} & r_{AB} & t_{AC} & t_{AD}\\ r_{BA} & r_{BB} & t_{BC} & t_{BD}\\ t_{CA} & t_{CB} & r_{CC} & r_{CD}\\ t_{DA} & t_{DB} & r_{DC} & r_{DD}\end{array}\right)
\end{equation}
and $r_{\alpha,\beta}$ ($t_{\alpha,\beta}$)  is the probability amplitude for reflection (transmission) of the electron from lead $\beta$ to lead $\alpha$. 
The creation operators are related by the hermitian conjugate of the $S$-matrix.
 
The field operators in the lead $\alpha$ are written as
\begin{equation}
\hat{\Psi}_\alpha(x,t)=\int dE e^{-iEt/\hbar} \left[\hat{a}_\alpha(E)e^{{\pm}ikx}+\sum_\beta s_{\alpha \beta}\hat{a}_\beta(E)e^{{\mp}ikx}\right] \left(\frac{m}{2 \pi \hbar^2 k}\right)^{\frac{1}{2}}
\end{equation}
\begin{equation}
\hat{\Psi}^\dagger_\alpha(x,t)=\int dE e^{iEt/\hbar} \left[\hat{a}^\dagger_\alpha(E)e^{{\mp}ikx}+\sum_\beta s^\dagger_{\alpha \beta}\hat{a}^\dagger_\beta(E)e^{{\pm}ikx}\right] \left(\frac{m}{2 \pi \hbar^2 k}\right)^{\frac{1}{2}}
\end{equation}       
where the upper (lower) sign on applies to leads on the left (right), $\beta=A,B,C,D$, $k=\sqrt{2mE}/\hbar$ is the wavevector of the electron in the lead, and  we have used the relationship between the incoming and outgoing creation and annihilation operators with $s_{\alpha \beta}$ representing the $S$-matrix element in row $\alpha$ and column $\beta$.  The normalization of the field operators conserves current in the system.  
We write the current operator in lead $\alpha$ as
\begin{equation}
\hat{I}_\alpha(x,t)=\frac{\hbar e}{2im} \left[\hat{\Psi}^\dagger_\alpha(x,t)\frac{\partial}{\partial x}\hat{\Psi}_\alpha(x,t)  -  {\biggl(} \frac{\partial}{\partial x}\hat{\Psi}^\dagger_\alpha(x,t) {\biggr)} \hat{\Psi}_\alpha(x,t)\right] .
\end{equation}

The correlation function of the currents in leads $\alpha$ and $\beta$ is defined as
\begin{equation}
C_{\alpha, \beta}(x,t;x',t')=\frac{1}{2}\left\langle \Delta\hat{I}_\alpha(x,t)\Delta\hat{I}_\beta(x',t')+\Delta\hat{I}_\beta(x',t')\Delta\hat{I}_\alpha(x,t) \right\rangle
\end{equation}
where $\Delta\hat{I}_\alpha = \hat{I}\alpha- \langle I_\alpha \rangle$ and the average ${\langle} {\rangle}$ is taken over all allowed electron energies.  The {\it{noise power}} (spectral density) , $S_{\alpha, \beta}(\omega)$, is then defined by the relation
\begin{equation}
2\pi\delta(\omega+\omega')S_{\alpha, \beta}(x,x';\omega)= \frac{1}{2}\left \langle \Delta\hat{I}_\alpha(x,\omega)\Delta\hat{I}_\beta(x',\omega')+\Delta\hat{I}_\beta(x',\omega')\Delta\hat{I}_\alpha(x,\omega) \right\rangle
\end{equation}
and is obtained by taking the Fourier transform of the time dependence in the correlation function $C_{\alpha,\beta}(x,t;x',t')$. 
The zero frequency {\it{shot noise}} is defined as $S_{\alpha, \beta}(0) \equiv {\displaystyle\lim_{\omega \to 0}}S_{\alpha,\beta}(x,x';\omega)$. In this limit, the spectral density becomes independent of the spatial variables $x$ and $x'$.

For our study of the $\sqrt{NOT}$ gate we compute the shot noise in the leads on the right of the scattering region given that a unit current enters one of the leads on the left of the scattering region.  Electrons in leads $A$ and $C$ correspond to the qubit in state $|1\rangle$, and electrons in leads $B$ and $D$ correspond to the qubit in state $|0\rangle$.  We will assume that the electrons enter the scattering region on the left in the state $|1\rangle$.  This requires a small bias voltage, $-V$, in reservoir $A$, so reservoir $A$ has a chemical potential ${\mu}_A={\mu}_0-eV$ and the other three reservoirs have chemical potentials ${\mu}_B={\mu}_C={\mu}_D={\mu}_0$.  We calculate the zero-frequency shot noise in lead $D$ and across leads $C$ and $D$ from the equation above and obtain
\begin{equation}
S_{D, D}(0) = {\displaystyle\frac{e^3 V}{h}{\rm coth}({\beta}eV/2)~\left( |t_{DA}|^2|t_{DB}|^2 + |t_{DA}|^2|r_{DC}|^2 + |t_{DA}|^2|r_{DD}|^2\right)}   
\end{equation}
and
\begin{equation}
S_{C ,D}(0) = {\displaystyle\frac{e^3 V}{h}}{\rm coth}({\beta}eV/2)~ {\rm Re}\left[t^{\dagger}_{CA}t_{CB}t^\dagger_{DB}t_{DA} + t^\dagger_{CA}r_{CC}r^\dagger_{DC}t_{DA} + t^\dagger_{CA}r_{CD}r^\dagger_{DD}t_{DA}\right]    
\end{equation}
where ${\rm Re}[~]$  denotes the real part of the quantity inside the brackets.

\section{Shot noise and fidelity of a $\sqrt{NOT}$ gate}

The $\sqrt{NOT}$ gate  takes a qubit initially in either state $|1\rangle$ or $|0\rangle$ into a linear superposition of the two with equal amplitude.  In their study of the stationary states of a $\sqrt{NOT}$ gate in electron waveguide qubits, Akguc et al. \cite{kn:akguc}  found a set of numerical parameters (spatial size, potential barrier height, incident energy) which allow for  a functioning $\sqrt{NOT}$ gate in a $GaAs/Al_xGa_{1-x}As$ based waveguide system.  The gate relied on a resonance of the scattering region to obtain a roughly equal distribution of electron probability in the $|1\rangle$ and $|0\rangle$ outgoing qubit states given an input of unit probability in the incoming $|1\rangle$ state.  For fixed spatial size and potential barrier height, they presented a plot of the outgoing electron probabilities in each of the four leads as a function of incoming electron momentum.  We have constructed a single parameter S-matrix which reproduces the qualitative behavior of the $\sqrt{NOT}$ gate found numerically in \cite{kn:akguc} for the $GaAs/Al_xGa_{1-x}As$ based waveguide system.  In this section, we use this single parameter S-matrix  to  find the fidelity of the output state as a function of the variable parameter and we calculate the zero-frequency shot noise in the output leads.

\begin{figure}[htb]
\begin{center}
\includegraphics{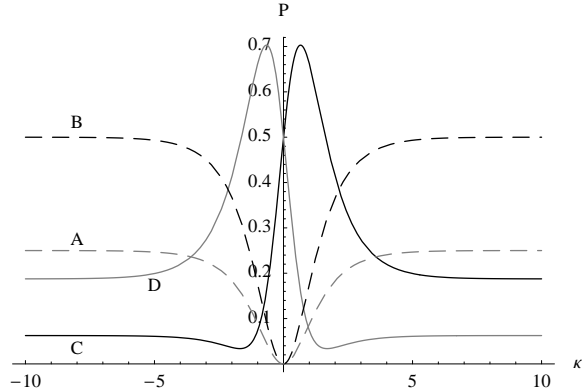}
\caption{A plot of the probability exiting leads $A$, $B$, $C$, and $D$ as a function of $\kappa$ given a unit current input in lead $A$.  At $\kappa=0$ a perfect $\sqrt{NOT}$ gate is obtained.}\label{probability}
\end{center}
\end{figure}

The single parameter  $S$-matrix is defined 
\begin{equation}
r_{AA}=-r_{BB}=-r_{CC}=r_{DD}=\sqrt{\frac{1}{4}\left[1-\mbox{sech}(\kappa)\right]}
\label{smat-1}
\end{equation}
\begin{equation}
-r_{BA}=-r_{AB}=r_{CD}=r_{DC}=\sqrt{\frac{1}{2}\left[1-\mbox{sech}(\kappa)\right]}
\end{equation}
\begin{equation}
-t_{AD}=t_{BC}=-t_{CB}=t_{DA}=\sqrt{\frac{5}{64}+\frac{1}{2}\left[\mbox{sech}(\kappa)+\frac{1}{8}\right]\left[\mbox{tanh}(\kappa)+\frac{3}{4}\right]}
\end{equation}
\begin{equation}
t_{AC}=t_{BD}=t_{CA}=-t_{DB}=\sqrt{\frac{5}{64}+\frac{1}{2}\left[\mbox{sech}(\kappa)+\frac{1}{8}\right]\left[\mbox{tanh}(-\kappa)+\frac{3}{4}\right]}
\label{smat-4}
\end{equation}
For an input of unit current in lead $A$, the probability exiting the various leads as a function of $\kappa$ is plotted in Fig. \ref{probability}.  When $\kappa=0$, half of the probability exits the $\sqrt{NOT}$ gate in lead $C$ and half in lead $D$.  This corresponds to the qubit state $(1/\sqrt{2})(|0\rangle + |1\rangle)$.  As $\kappa$ differs from zero the transmission properties of the gate change, preventing the output of a perfect superposition state.

\begin{figure}[htb]
\begin{center}
\includegraphics{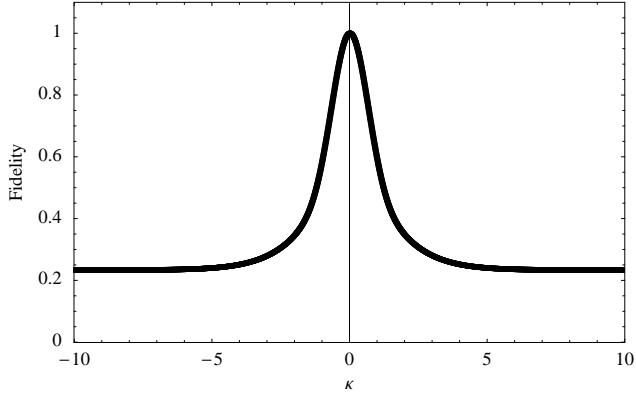}
\caption{The fidelity of the $\sqrt{NOT}$ gate as a function of $\kappa$.  The fidelity is unity when $\kappa=0$.  }\label{fidelity}
\end{center}
\end{figure}

We define the fidelity of the gate as the absolute value squared of the overlap of the output state $|\Phi_0{\rangle} = (c_A, c_B, c_C, c_D)^T$ with the perfect superposition state $|\Xi{\rangle}= (0, 0, 1/\sqrt{2}, 1/\sqrt{2})^T$,
\begin{equation}
F=|{\langle}\Phi^\dagger_0~|\Xi{\rangle}|^2
\end{equation}
The fidelity of the $\sqrt{NOT}$ gate is plotted as a function of $\kappa$ in Fig. \ref{fidelity}.  The gate has a maximum fidelity at $\kappa=0$ when a perfect superposition state is obtained.  

\begin{figure}[htb]
\begin{center}
\includegraphics{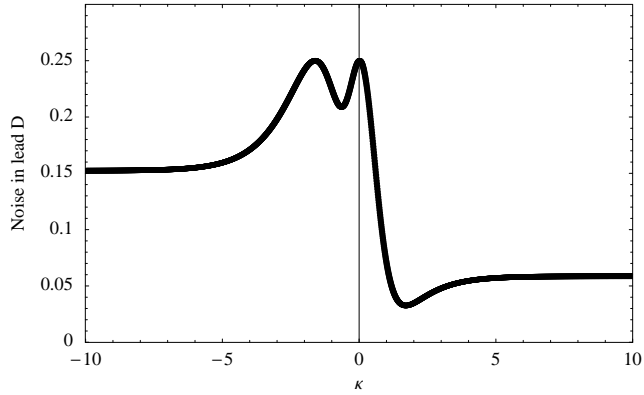}
\caption{The shot noise in lead $D$ in fractions of $(e^3V/h){\rm coth}(\beta eV/2)$ is plotted as a function of $\kappa$.  We see that the shot noise has two maximums, one before the gate is on resonance, and one on resonance when $\kappa=0$.  This is due to the fact that the noise is greatest when the probability of exiting lead $D$ is half, which occurs twice between $-10 < \kappa < 10$.   }\label{noiseDD}
\end{center}
\end{figure}

\begin{figure}[htb]
\begin{center}
\includegraphics{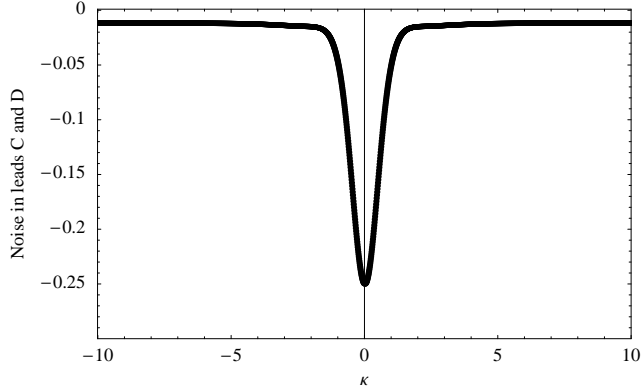}
\caption{The shot noise across leads $C$ and $D$ in fractions of $(e^3V/h){\rm coth}(\beta eV/2)$ is plotted as a function of $\kappa$.  Here the shot noise has a single minimum when the gate is on resonance at $\kappa=0$.  The magnitude of the noise is greatest when the probability of exiting both leads $C$ and $D$ is half, which occurs only at $\kappa=0$. }\label{noiseCD}
\end{center}
\end{figure}

We now look at the noise characteristics of the $\sqrt{NOT}$ gate described by our single parameter $S$-matrix defined in Eqs. (\ref{smat-1}) - (\ref{smat-4}).  In Fig. \ref{noiseDD} and Fig. \ref{noiseCD} we plot the zero-frequency shot noise in lead $D$ and across leads $C$ and $D$ respectively.  The noise is plotted in fractions of $(e^3V/h){\rm coth}(\beta eV/2)$ for simplicity.    The noise in lead $D$ has two maximums, one for negative $\kappa$ and another when  $\kappa=0$.  The magnitude of the shot noise is largest when the probability of exiting in lead $D$ is equal to $1/2$.  This occurs twice between $-10 < \kappa < 10$.  
The noise across leads $C$ and $D$ has a single minimum which occurs when $\kappa=0$.  The magnitude of the shot noise across the leads is greatest when the probability of exiting  the gate in both leads $C$ and $D$ is $1/2$, which occurs only once at $\kappa=0$.

Shot noise measurements in a single lead do not allow the determination of the fidelity of a $\sqrt{NOT}$ gate uniquely.  The noise in the single lead will be at a maximum any time there is a probability of $1/2$ that an electron will exit the gate in that lead, but this says nothing about the probability that an electron will exit any of the other leads.  Shot noise measurements across the two leads on the output side of the device do, however, allow for the unique determination of the fidelity of a $\sqrt{NOT}$ gate.  In a physical system it may be useful to measure the fidelity of the gate in this manner, as it is possible to have an equal splitting of the current through such a gate but also contain some small amount of reflection back into the originating lead.  Shot noise measurements may then be able to differentiate a gate which simply splits an incoming current equally from a gate which behaves as a high fidelity $\sqrt{NOT}$ gate.

\section{Conclusion}

The construction of high-fidelity $\sqrt{NOT}$ gates in ballistic electron waveguide networks is essential to  the eventual implementation of the quantum algorithms.  The reflection of electron probability at the $\sqrt{NOT}$ gates causes a rapid decay of overall computation fidelity.  Measurements of the shot noise across leads on the output side of a gate can describe the fidelity characteristics of the gate.  Shot noise measurements can be used to discriminate between an electron waveguide gate that simply splits incoming current, and one which is of high-fidelity with very little reflection back towards the input side of the device.  

\section{Acknowledgments}

The authors thank the Robert A. Welch Foundation (Grant No. F-1051)
and the Engineering Research Program of the Office of Basic Energy
Sciences at the U.S. Department of Energy (Grant No.
DE-FG03-94ER14465) for support of this work. Author LER thanks the 
Office of Naval Research (Grant No. N00014-03-1-0639) for partial support of this work.
The authors also thank Tianyi Yang for useful discussions.

\pagebreak

\end{document}